\documentclass[aps,pre,preprint,endfloats, showpacs]{revtex4}
\usepackage{graphicx,bm,amssymb}
\newcommand{\omv}{\mbox{\boldmath$\omega$}}

\preprint{LA-UR 08-8194}

\begin{document}

\title{Cascade time-scales for energy and helicity 
in homogeneous, isotropic turbulence}
\author{Susan Kurien}
\affiliation{Center for Nonlinear Studies and Theoretical Division,~Los
  Alamos National Laboratory, Los Alamos, New Mexico 87545, U.S.A.}
\author{Mark A. Taylor}
\affiliation{Computer and Computational Sciences Division, Los Alamos National
  Laboratory, Los Alamos, New Mexico 87545, U.S.A.}
\author{Takeshi Matsumoto}
\affiliation{Department of Physics, Kyoto University, Kitashirakawa Oiwakecho
  Sakyo-ku, Kyoto 606-8502, Japan}


\date{\today}
\begin{abstract}
  We extend the Kolmogorov phenomenology for the scaling of energy
  spectra in high-Reynolds number turbulence, to explicitly include
  the effect of helicity. There exists a time-scale $\tau_H$ for
  helicity transfer in homogeneous, isotropic turbulence with
  helicity. We arrive at this timescale using the phenomenological
  arguments used by Kraichnan to derive the timescale $\tau_E$ for
  energy transfer (J.  Fluid Mech.~{\bf 47},~525--535~(1971)). We show
  that in general $\tau_H$ may not be neglected compared to $\tau_E$,
  even for rather low relative helicity. We then deduce an inertial
  range joint cascade of energy and helicity in which the dynamics are
  dominated by $\tau_E$ in the low wavenumbers with both energy and
  helicity spectra scaling as $k^{-5/3}$; and by $\tau_H$ at larger
  wavenumbers with spectra scaling as $k^{-4/3}$. We demonstrate how,
  within this phenomenology, the commonly observed ``bottleneck'' in
  the energy spectrum might be explained.  We derive a wavenumber
  $k_h$ which is less than the Kolmogorov dissipation wavenumber, at
  which both energy and helicity cascades terminate due to dissipation
  effects. Data from direct numerical simulations are used to check
  our predictions.


\end{abstract}
\pacs{47.27.Gs,47.27.Jv,47.27.Eq}
\maketitle

Energy and helicity \cite{Moreau,Moffat} are the two known inviscid invariants
of the Navier--Stokes equations. It was postulated in \cite{BFLLM73} that
in isotropic flows with helicity, these quantities cascade
together from large to small scales. This joint forward cascade of energy and helicity has been verified by direct numerical simulations, most recently at a 
resolution of $512^3$ gridpoints \cite{CCE03}.
Kraichnan
\cite{Kraichnan71} defined the shear time-scale $\tau_E$ for energy transfer,
 based solely on energy dynamics. Assuming that
helicity dynamics are also controlled by $\tau_E$, a $k^{-5/3}$
inertial range scaling was established for both energy and helicity
spectra \cite{BFLLM73}.

\begin{figure}
\includegraphics[scale=0.45]{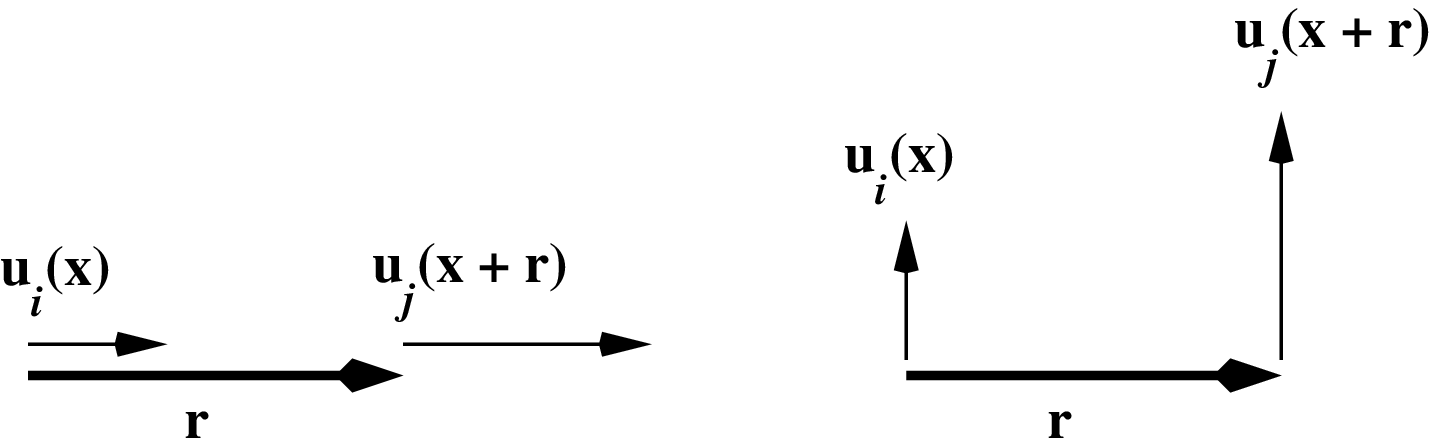}
\caption{The in-plane longitudinal and transverse correlation
  configurations which contribute to the isotropic symmetric
  correlation function $R^S_{ij}({\bf r})$. \label{symcor}}
\vspace{0.2in}
\includegraphics[scale=0.45]{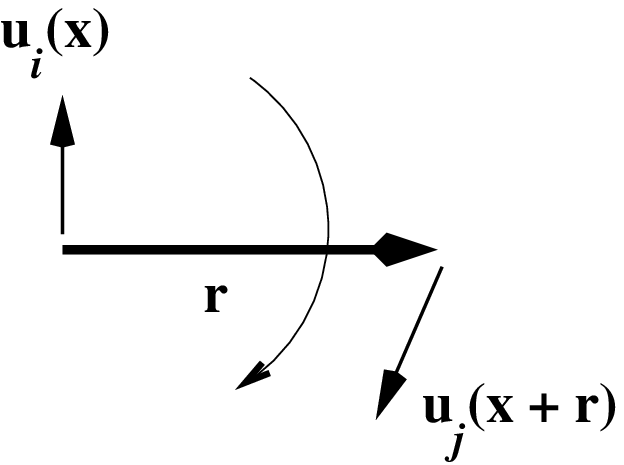}
\caption{The out-of-plane correlation
  configuration which contributes to the isotropic antisymmetric
  correlation function $R^A_{ij}({\bf r})$. The intrinsic `handedness'
  of this configuration, indicated by the curved arrow, cannot
  appear in the geometry of Fig.~\ref{symcor}. \label{asymcor}}
\end{figure}

We would first like to ascribe spatial geometrical properties to the
types of quantities used to derive the relevant timescales.  We recall
the spectral formulation $\langle \tilde u_i({\bf k})\tilde u_j^*({\bf
  k}) \rangle$ of the two-point velocity correlation function in
isotropic, homogeneous, statistically stationary turbulence.  It may
be decomposed into its index-symmetric and index-antisymmetric parts
as
\begin{eqnarray}
E_{ij}({\bf k}) &=& \frac{1}{2}\Big(\langle \tilde{u}_i({\bf k})\tilde{u}_j^*({\bf k}) \rangle
+ \langle \tilde{u}_j({\bf k})\tilde{u}_i^*({\bf k}) \rangle \Big),\label{spec}\\
\widetilde E_{ij}({\bf k}) &=& \frac{1}{2}\Big(\langle \tilde{u}_i({\bf
  k})\tilde{u}_j^*({\bf k}) \rangle - \langle \tilde{u}_j({\bf k})\tilde{u}_i^*({\bf k})
\rangle \Big),
\label{cospec}
\end{eqnarray}
where ${\bf \tilde u}_i = {\tilde u}_i {\bf \hat i}$ and $\tilde u_i$ is the magnitude of the
$i$th component of the velocity vector in a chosen cartesian coordinate 
system. Eq.~(\ref{spec}) when contracted with the
projection operator $\delta_{ij}/2$ and then averaged over ${\bf \hat
  k}$ gives the energy spectrum $E(k)$. It is therefore clear that the
types of correlations contributing to $E(k)$ are those in which $i=j$
and hence ${\bf \hat i}$, ${\bf \hat j}$ and the
unit wavevector ${\hat {\bf k}}$ all lie in
the same plane. The corresponding picture in real space is to consider
the index-symmetric two-point spatial correlation functions
$R^S_{ij}({\bf r}) = \frac{1}{2}\langle u_i({\bf x})u_j({\bf x}+{\bf
  r}) + u_j({\bf x})u_i({\bf }x+{\bf r})\rangle$ which has the tensor
representation $\sim A(r)\delta_{ij} + B(r)\frac{r_i r_j}{r^2}$ for
the isotropic case; the incompressibility constraint gives a
relationship between $A(r)$ and $B(r)$. This index-symmetric
correlation function thus has non-zero contributions when ${\hat {\bf i}}$, ${\hat {\bf j}}$ and ${\hat {\bf r}}$ are co-planar. 
We will refer to these as 'in-plane' correlations (see
Fig.~\ref{symcor} for a sketch of these types of isotropic
correlations). Similarly, Eq.~(\ref{cospec}) when contracted with the
antisymmetric curl operator $\widehat i\varepsilon_{ijl}{k_l}$, where
$\widehat i = \sqrt{-1}$, and then averaged over ${\bf {\hat k}}$
gives the total helicity density $H(k) = 2 k \widetilde E(k)$. (Note
that this relationship is distinct from the Schwartz-inequality
$|H(k)| \le 2 k E(k)$). Therefore, the types of correlations
contributing to $\widetilde E_{ij}(k)$ (and hence to $H(k)$) are those
in which ${\bf \hat i}$, ${\bf \hat j}$ and unit wavevector ${\bf
  {\hat k}}$ are mutually
orthogonal.  Again, the corresponding formulation in real space is the
index-antisymmetric two-point spatial correlation functions
$R^A_{ij}({\bf r}) = \frac{1}{2}\langle u_i({\bf x})u_j({\bf x}+{\bf r}) -
u_j({\bf x})u_i({\bf x}+{\bf r})\rangle$ which has the tensor
representation $\sim \varepsilon_{ijl}{r_l/r}$ and thus has non-zero
contributions when ${\bf \hat i}$, ${\bf \hat j}$ and ${\hat {\bf r}}$ are mutually
orthogonal to each other (see Fig.~\ref{asymcor} for a sketch).  We
will refer to these as 'out-of-plane' correlations. Here $E(k) =
\sum_{|{\bf k}| = k}\frac{1}{2}|\tilde {\bf u}({\bf k})|^2$ and $H(k)
= \sum_{|{\bf k}| = k} \tilde {\bf u}({\bf k})\cdot \tilde{\omv}(-{\bf
  k})$ where $\tilde {\bf u}({\bf k})$ and $\tilde{\omv}({\bf k})$ are
the fourier transforms of the velocity ${\bf u}({\bf x})$ and the
vorticity $\omv({\bf x}) = \nabla \times {\bf u}({\bf x})$
respectively.

The Kraichnan time-scale for energy transfer, $\tau_E$, corresponds to
correlations of the type $E_{ij}({\bf k})$ (Eq. (\ref{spec})) which
arise due to shearing motions in the plane of coordinates $i$, $j$ and
unit wavenumber $\hat {\bf k}$ \cite{Kraichnan71}. Such in-plane
shearing motions cannot give rise to correlations of the type
$\widetilde E_{ij}({\bf k})$ (Eq.  (\ref{cospec})) which relate
orthogonal components $u_i {\bf\hat i}$ and $u_j {\bf \hat j}$ across
the third mutually orthogonal direction $\hat {\bf k}$. For this we
require out-of-plane shearing motions as depicted in
Fig.~\ref{asymcor}, which are provided by the presence of helicity
\cite{Betchov61,Kurien03}. We first derive the time-scale, $\tau_H$,
associated with such an out-of-plane shear. The governing factor is
the relative helicity $|H(k)| /(2 k E(k))$ which will be shown to fall
off linearly in wavenumber restoring parity as $k$ becomes very large.
Crucially, we will show that the ratio $\tau_E/\tau_H \sim (|H(k)| /(2
k E(k)))^{1/2}$, which decays slower than the relative helicity.
Therefore, the effect of $\tau_H$ cannot be neglected. We demonstrate
the effect of this new timescale on energy and helicity spectra, and
offer an interpretation of the ``bottleneck'' effect observed in
measured energy spectra. Finally, the new dynamics reveal a
dissipation scale which is larger than the Kolmogorov dissipation
scale, suggesting that the joint cascade is truncated sooner in
wavenumber space if helicity is present.

We performed two simulations of the three-dimensional (3-$d$), forced
Navier-Stokes equation in a unit-periodic box with $512$ (data I), and
$1024$ (data II) grid points to a side respectively. In these units,
the wavenumber $k$ is in integer multiples of $2\pi$. Energy and
helicity were injected into the flow for $ k \le 2$ at each time-step.
The forcing scheme was the same as in \cite{TayKurEyi03}. For case I
we imposed maximum helicity in $k \le 2$ \cite{PolShtil89,CCE03},
resulting in a mean helicity over time of -26.8 in the units of our
simulation.  For case II the helicity input was uncontrolled and
random, resulting in a mean helicity of -0.12 which is essentially
zero compared to case I. The spectra for case I
were averaged over 40 snapshots spanning 8 large-eddy turnover times
after spin-up. The spectra for case II were averaged over 48 snapshots
spanning 2 large-eddy turnover times after spin-up. The spin-up time
in each case was defined to be when the input rate of energy matched
the dissipation rate of energy, the flow having achieved statistically
steady state.  Additional parameters of the simulations are given in
Table \ref{params}.

\begin{table}
\begin{tabular}{|c|c|c|c|c|c|c|c|c|c|}
\hline
&$N$ & $\nu \times 10^{4}$& $R_\lambda$ &$E$ & $\varepsilon$ & $H$
& $h$  & $\eta_\varepsilon \times 10^{3} $& $\eta_h \times 10^{4} $   \\
I&512 & 1& 270& 1.72 & 1.51 & -26.8 & 62.2 & 1.7 & 9 \\
II& 1024& 0.35 & 430 & 1.87 & 1.75 &-0.12 & 13.2 & 1.3 &4  \\
\hline
\end{tabular}
\caption{Parameters of the numerical simulations I and II. $\nu$ - viscosity; $R_\lambda$ -
  Taylor Reynolds number; mean total energy $E = \frac{1}{2}\sum_k \tilde {\bf
  u}({\bf k})^2$; $\varepsilon$ - mean energy dissipation rate; 
mean total helicity $H = \sum_k \tilde {\bf u}({\bf k}) \cdot \tilde {\omv}(-{\bf k})$; $h$ - mean helicity dissipation rate; $\eta_\varepsilon =
  (\nu^3/\varepsilon)^{1/4}$; $\eta_h = (\nu^3/h)^{1/5}$.  
\label{params}}
\end{table}



We recall the introduction in \cite{Kraichnan71} of the distortion timescale
(or eddy-turnover time) of an eddy with wavenumber $k$
\begin{equation}
\tau_E^2 \sim \Big (\int_0^k E(p) p^2 dp \Big)^{-1} \sim \Big(E(k) k^3\Big)^{-1}.
\label{tauE}
\end{equation}
where Kraichnan asssumes that only wavenumbers $\lesssim k$ will have a
shearing action on wavenumbers of order $k$; the effects from
wavenumbers $> k$ will average out. Notice that the local shear
timescale thus defined depends on the in-plane correlations which
contribute to the energy spectrum as described above.  Analogously, we
can define the time-scale $\tau_H$ for out-of-plane distortions of an
eddy, from the antisymmetric co-spectrum,
\begin{eqnarray}
\tau_H^2 \sim \Big (\int_0^k |\widetilde E(p)|p^2 dp\Big)^{-1} &\sim&
\Big(\frac{1}{2}|H(k)|k^2\Big)^{-1}.
\label{tauH}
\end{eqnarray}
The distortion or shear corresponding to $\tau_H$ is different from
the distortion corresponding to $\tau_E$ of \cite{Kraichnan71}. The
shear in the former case derives from the out-of-plane correlations
contributing to the antisymmetric co-spectrum and hence to the
helicity spectrum. The shear motion in this case is an out-of-plane
twist for which the time-scale is different than the in-plane shear
rate corresponding to $\tau_E$.

We estimate the transfer rate (flux) of helicity through wavenumber
$k$, $F_H(k) \sim {k |H(k)|}/{\tau_H} \sim {k^2 |H(k)|^{3/2}}$. Assuming steady-state, inertial range behavior with constant
flux of helicity $F_H(k) = h$, the mean helicity dissipation rate, we
obtain
\begin{eqnarray}
H(k) &\approx&  h^{2/3} k^{-4/3}.
\label{H43}
\end{eqnarray}
We compare $\tau_H$ with the time-scale for energy transfer
of Eq.~(\ref{tauE}),
\begin{equation}
\frac{\tau_E}{\tau_H} \sim \Big(\frac{|H(k)|}{2 k E(k)}\Big)^{1/2}.
\label{tau_ratio}
\end{equation}
Since $[~|H(k)|/(2 k E(k))~]^{1/2} \gg |H(k)|/(2 k E(k))$ as the
latter tends to zero, Eq.~(\ref{tau_ratio}) implies that even for
small values of the relative helicity, the time-scales can become
comparable. This is a fundamental point of difference from previous
works in which the presence of helicity was considered inconsequential
\cite{BFLLM73}. In previous arguments the fact that relative helicity
must go to zero
as $1/k$ in wavenumber space meant that helicity
could not have an effect on the long term dynamics since it must
eventually be dominated by energy, restoring parity. Our present
analysis shows that while the relative helicity does indeed go rapidly
to zero, the relative $timescale$ of helicity and energy transfer
vanishes much slower. In other words, while the energy timescale is
always faster, because of the Schwartz equality, the helicity timescale
can remain comparable to it well into the large wavenumbers. (For the
initial value problems (decaying turbulence), the evolution
 equation of the relative helicity $H(k)/(2kE(k))$ and its analytical
 bounds can be found in \cite{ConMaj88}.)

With this second timescale at hand, we can now justifiably ask what
the effect of $\tau_H$ will be on the energy spectrum. The energy flux
$\varepsilon$ through wavenumber $k$ is
\begin{eqnarray}
\varepsilon &\sim& \frac{kE(k)}{\tau_H} \sim {k E(k)} H(k)^{1/2} k.
\label{eflux}
\end{eqnarray}
Using Eq. (\ref{H43}) in Eq. (\ref{eflux}), we get 
\begin{equation}
E(k)\approx \varepsilon h^{-1/3}k^{-4/3}. 
\label{espec_hel}
\end{equation}
To summarize thus far, the $\tau_E$ dynamics result in Kolmogorov
$k^{-5/3}$ scaling in both energy and helicity spectra, whereas
$\tau_H$ dynamics result in $k^{-4/3}$ scaling in both. Clearly, the
steeper $k^{-5/3}$ scaling should dominate in the low-wavenumbers
while the $k^{-4/3}$ should manifest in the higher wavenumbers. We
emphasize that in order for the latter scaling to be visible in the
high wavenumbers, $\tau_H$ cannot be too much slower than $\tau_E$. As
shown above, according to Eq.~(\ref{tau_ratio}), this may occur for
very modest relative helicity in the high wavenumbers, contrary to
previous assumptions.  The main point of our paper
is that in
general, the helicity timescale $\tau_H$ may not be ignored.

\begin{figure}
\includegraphics[scale = 0.35]{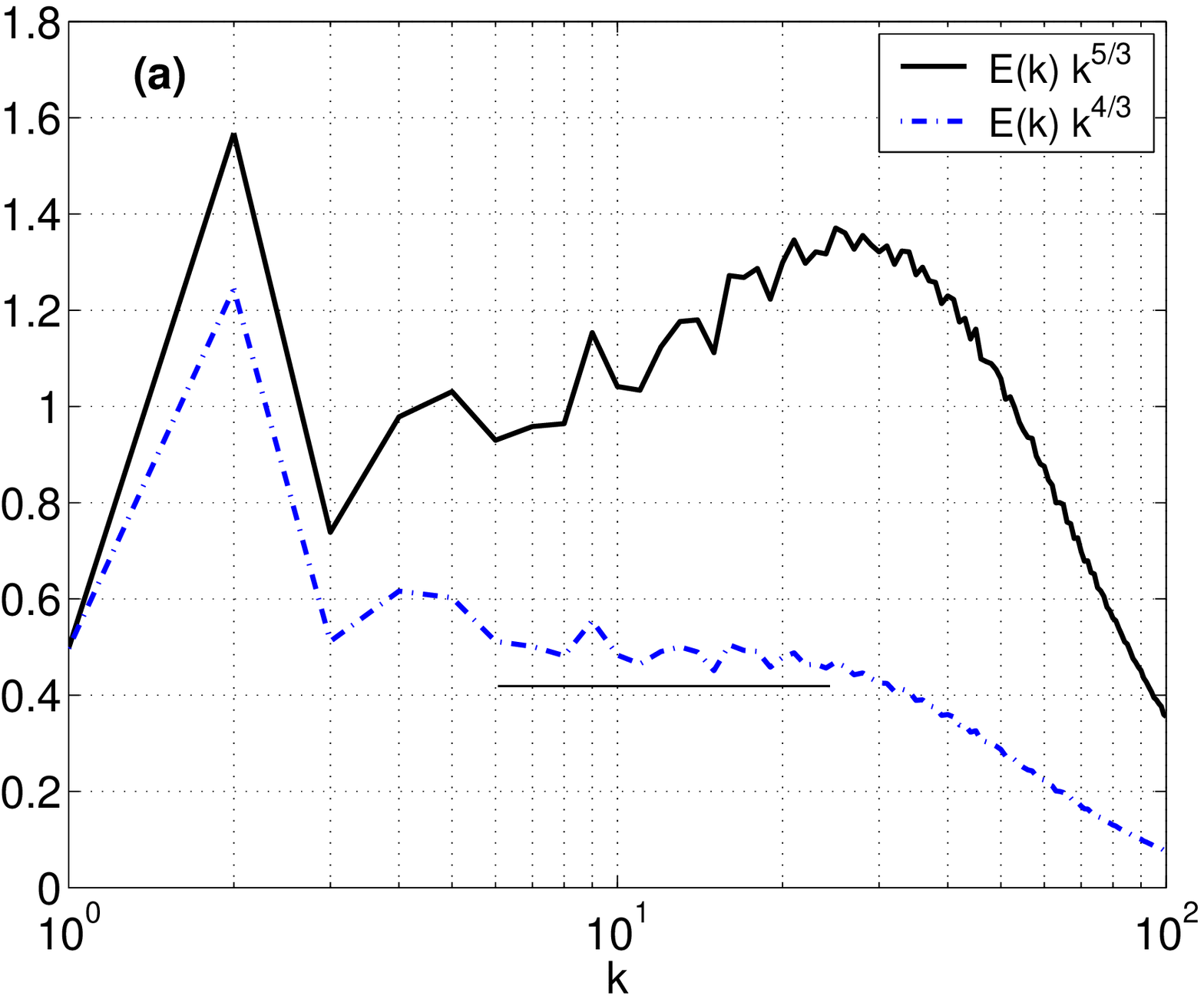}
\includegraphics[scale = 0.35]{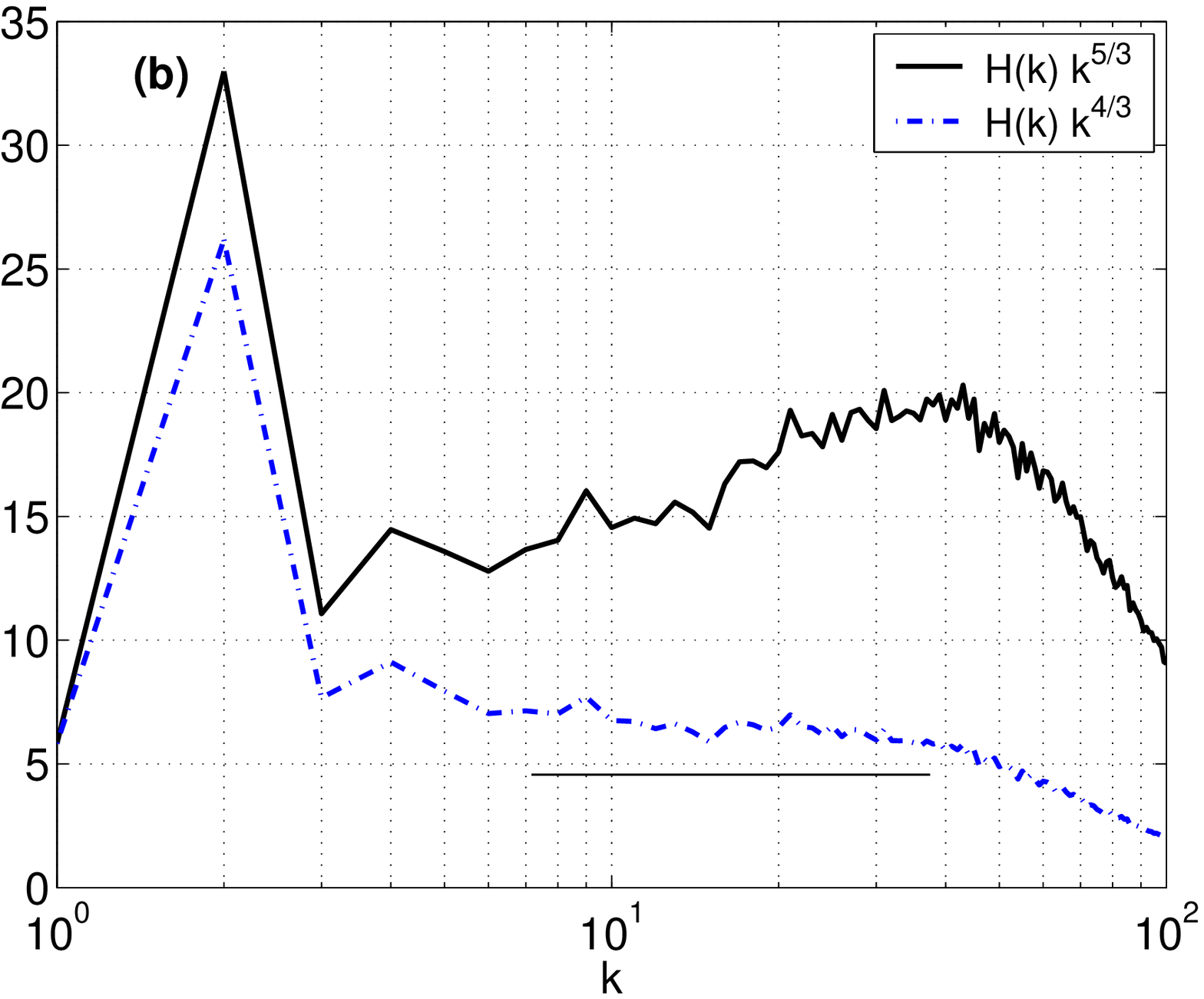}
\includegraphics[scale = 0.35]{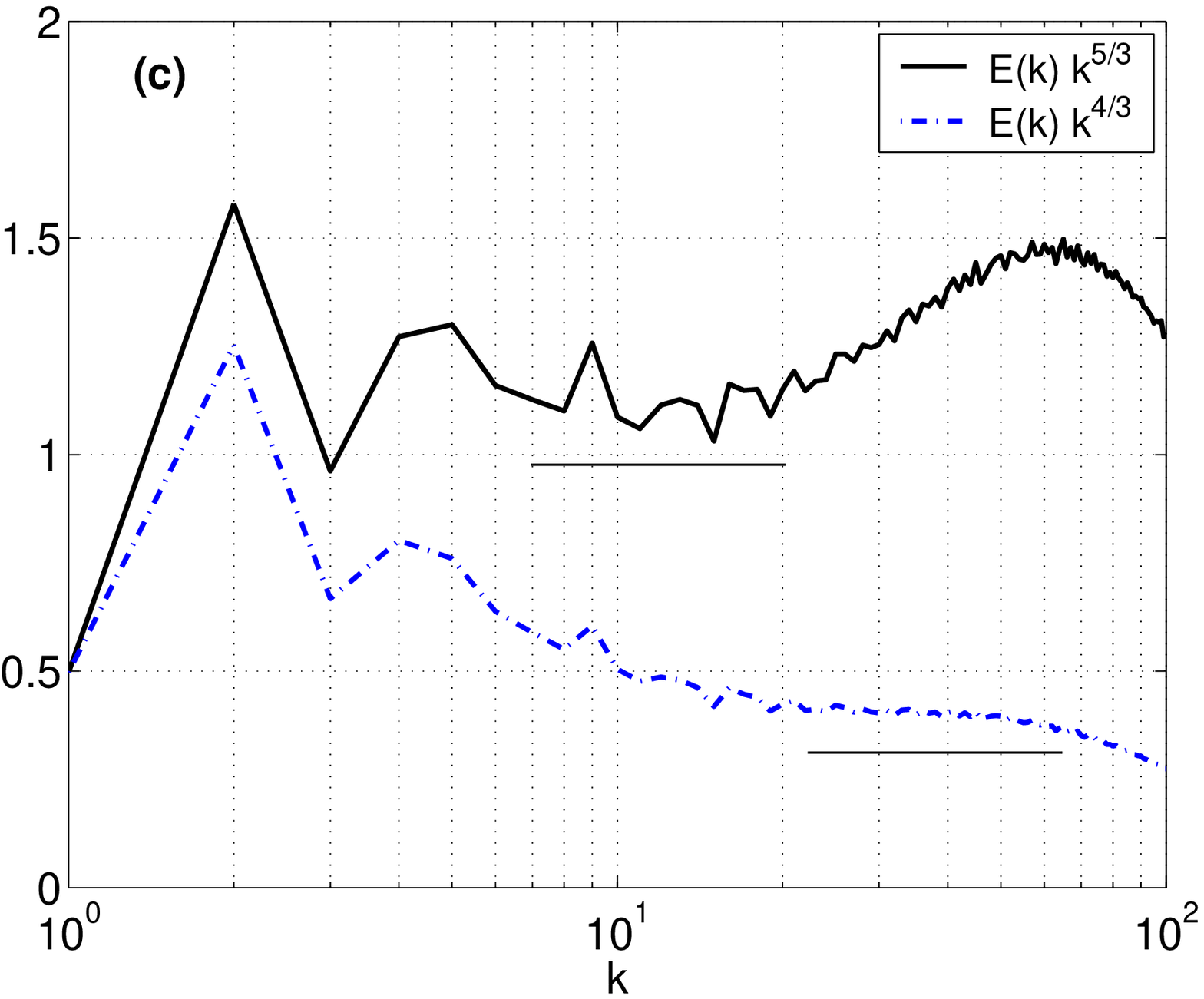}
\caption{Compensated spectra of (a) energy and (b) helicity,  
  for data I. Observe the $k^{-4/3}$ scaling (dashed curve)
  range indicated by the horizontal line segment.
  (c) Compensated energy spectrum for data II. The horizontal line
  segments indicate ranges of $k^{-5/3}$ (solid curve) and $k^{-4/3}$
  (dashed curve) scaling.
  \label{spectra}}
\end{figure}

\begin{figure}
\includegraphics[scale = 0.35]{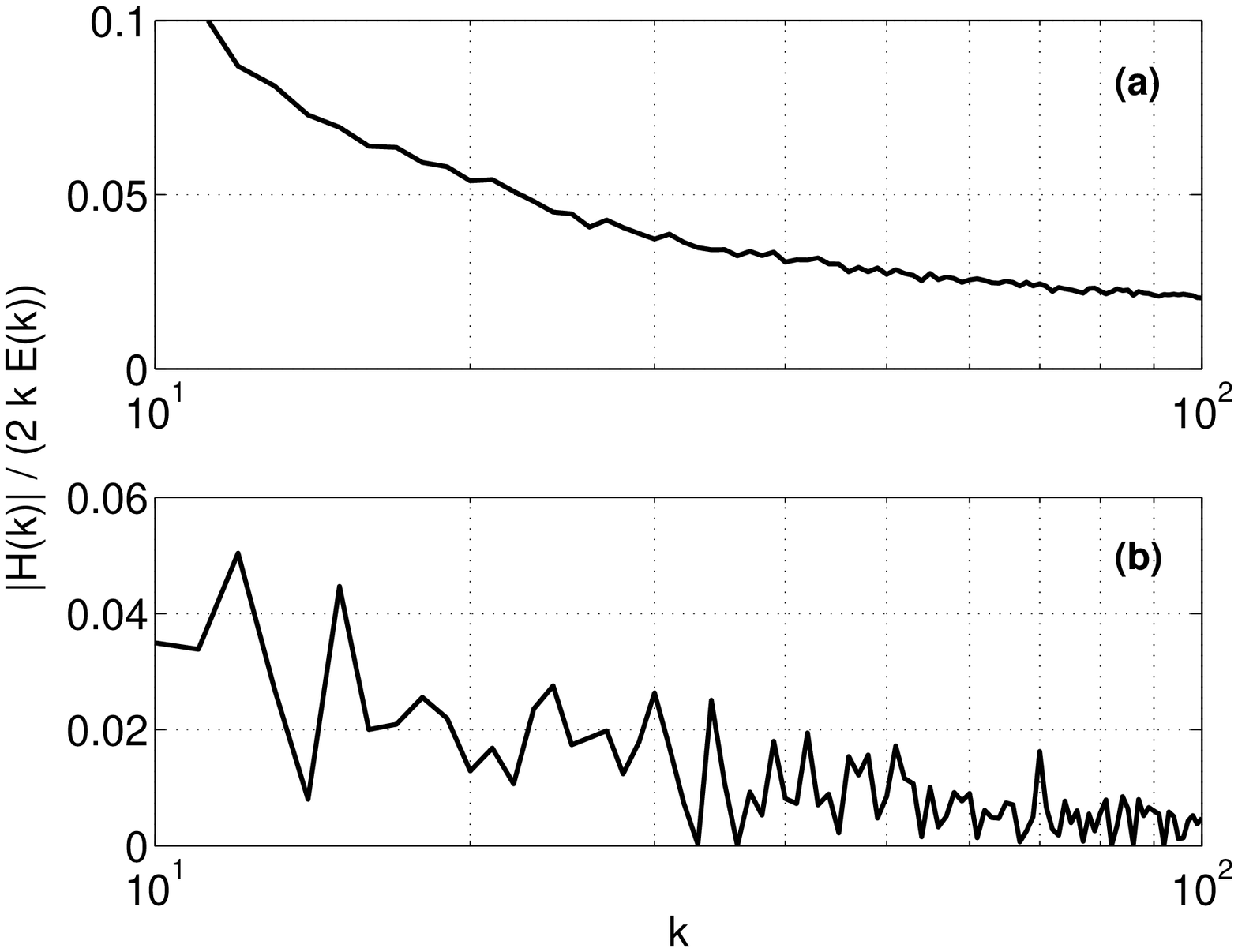}
\caption{The relative helicities for (a) data I and (b)
  data II, in the range $10 < k < 100$, where Fig.~\ref{spectra} shows
  $k^{-4/3}$ scaling. The relative helicity lies between 3\% and 10\%
  in (a) and between 1\% and 5\% in (b). The corresponding ratio
  of timescales $\tau_E/\tau_H$ (Eq.~(\ref{tau_ratio})), is estimated
  to be upto 32\% for data I upto 22\% for data II.\label{relh1024} }
\end{figure}

Figure \ref{spectra} shows the energy and helicity spectra from our
simulations. Each of the spectra are compensated by $k^{4/3}$ and by
$k^{5/3}$ in order to distinguish the dominant scaling. For the
strongly helical case I, there is good agreement with $k^{-4/3}$
scaling for both the energy and helicity spectra in approximately the
same range (Figures \ref{spectra}(a) and (b)) of slightly less than a
decade. Note that the compensation with $k^{5/3}$ results in the
commonly observed `bottleneck' phenomenon which we will discuss below.
The energy spectrum of II (Fig.~\ref{spectra}(c)) shows a range of
$k^{-5/3}$ scaling followed by a range of $k^{-4/3}$ scaling (which
appears as a bottleneck in the $k^{-5/3}$ compensated plot). The
scaling ranges are modest even at this high resolution of $1024^3$, but
nonetheless the results are telling; the scaling is most certainly not
$k^{-5/3}$ throughout and the agreement with $k^{-4/3}$ though over a
short range, certainly indicates shallower than $k^{-5/3}$ scaling
behavior of the bottleneck region. The relative helicities in the
range $10 < k < 100$, where the $k^{-4/3}$ scaling is seen, are shown
in Fig. \ref{relh1024}. For I, the relative helicity falls from about
10\% to about 3\% corresponding to $\tau_E/\tau_H$ ranging from 32\%
to 17\% according to Eq.  (\ref{tau_ratio}). Despite the negligibly
small total helicity $H = -0.12$ of II, and its noisy helicity
spectrum (not shown), its relative helicity values lie between $1\%$
and $5\%$. This implies that $\tau_E/\tau_H$ could be as much as
$22\%$. In both cases $\tau_H$ might in fact not be much longer than
$\tau_E$. It is important at this stage to comment on the appearance
of a helicity-dependent scaling feature in flow II which is nominally
helicity-free on average. The first point is that zero average
helicity does not imply that the average helicity spectrum $H(k)$ is
zero for all $k$. In fact, we only input energy and helicity in the
low-wavenumbers, the Navier-Stokes dynamics determines the helicity in
all other wavenumber, including the highest wavenumbers where in fact
there is the well-known viscous helicity production. There is
therefore no control of the helicity in an given wavenumber and the
spectrum is generally not zero everywhere. The second point is that
our analysis shows that it is not the total helicity but the
$relative$ helicity which determines the trade-off between the two
timescales. Given these two points it is not contradictory to measure
$k^{-4/3}$ spectral scaling in the flow with negligible mean helicity.
In fact, this flow is probably more similar to most experimental flows
which are close to helicity-free in the mean but with uncontrolled and
often unknown helicity spectra \cite{KhoShaTsi01}.

These results are the first indication of the possibility of
$k^{-4/3}$ scaling ranges in both energy and helicity spectra
simultaneously. The possibility of a `pure' or 'maximal' forward
cascade of helicity scaling as $k^{-4/3}$ \cite{BFLLM73}, with inverse
cascade of energy scaling as $k^{-7/3}$ does not arise.  This is
because in our analysis we have retained the effect of the helical
time-scale $\tau_H$ and allowed it to modify the spectral dynamics.
Since the scaling corresponding to $\tau_H$ is $k^{-4/3}$, a slower
decay than $k^{-5/3}$, its `signature' in the spectra can dominate at
large $k$ even as the overall parity is being restored.

Based on the analysis above, we propose that the bottleneck in the
total energy spectrum is in fact a {\it change in the scaling} of the
energy spectrum, from a $k^{-5/3}$ regime in which the $\tau_E$
dynamics dominate to a less steep $k^{-4/3}$ regime in which the
$\tau_H$ dynamics become significant. We will use the kinematic
arguments of \cite{DHYB03} with our new phenomenology and dynamics to
analyse the bottleneck for such a helical
influence.  In simulations, it is possible to compute the total energy
spectrum $E(k) = (1/2)\sum_{|{\bf k}|=k}|{\bf \tilde u}({\bf k})|^2$
as a sum over a shell of radius $k$ rather accurately. In experiments 
it is convenient to measure the one-dimensional (1-$d$)
longitudinal and transverse spectra along the measurement
direction, say $z$. In our 3-$d$ flow simulation we calculate the
1-$d$ spectra as follows.  The 1-$d$ fourier transform in the
$z$-direction of the velocity ${\bf u}({\bf x})$ is ${\bf {\tilde
    {\tilde u}}}(x,y,k_z) = (1/N)\sum^{N}_{n=1}
e^{ik_z z_n}{\bf u}(x,y,z_n)$ where $0\le k_z \le \pi/\delta z$. The
longitudinal (transverse) 1-$d$ spectrum, averaged over the $x$-$y$
plane, is defined by
\begin{equation}
E_{L(T)}(k_z) = \frac{1}{(2N^2)}\sum^N_{p,q=1}|{\tilde {\tilde
    u}}_{z(\perp)}(x_p,y_q,k_z)|^2,
\label{spec1d}
\end{equation}  
where $|{\tilde {\tilde u}}_\perp|^2 = |{\tilde {\tilde u}}_x|^2 +
|{\tilde {\tilde u}}_y|^2$.
In isotropic flow, the 1-$d$ spectra
should be independent of the direction in which the fourier transform
is performed. Our time-averaged longitudinal spectra computed in the the
$x$, $y$ and $z$ directions all collapse and the transverse spectra do
the same. We average the 1-$d$ spectra in the
three coordinate directions and drop the use of the subscript $z$ to
denote the direction of the fourier transform. In isotropic flow
there is a relation between the 1-$d$ and 3-$d$ spectra
\cite{MYII,DHYB03},
\begin{equation}
E(k) = -k\Big (\frac{dE_L}{dk} + 2 \frac{dE_T}{dk}\Big).
\label{local}
\end{equation} 
Our spectra satisfy Eq.~(\ref{local}) very well for $k\geq 10$ and
fairly well in the lower wavenumbers. As emphasized in \cite{DHYB03},
Eq. (\ref{local}) is a local relationship, wherein the functional form
of the total spectrum $E(k)$ is fully determined by the local behavior
of $E_L$ and $E_T$ at a given wave number.  For homogeneous, isotropic
flows with helicity, it is reasonable to suppose that $E_L(k)$ (see
definition in Eq.~(\ref{spec1d})) would mainly carry contributions
from the in-plane shear time-scale $\tau_E$. However $E_T(k)$, which
is related to the transverse components of the velocity fourier
transform, could be influenced by $\tau_H$ dynamics coming from
$\widetilde E(k)$. The correlation time between transverse components
can be slowed down by the dynamics of $\widetilde E(k)$ which arise
due to the presence of helicity. Such coupling may not be deduced from
kinematic arguments, it requires proper consideration of the new
dynamics.  Furthermore, it is not possible to see this coupling in the
unclosed lowest-order K\'arm\'an-Howarth dyanamical equations wherein
the symmetric and antisymmetric parts completely decouple for
homogeneous flows \cite{Kurien03}. It was pointed out in
\cite{Betchov61} that higher-order statistical equations for helical
turbulence should show the coupling.  The coupling is seen explicitly
in the EDQNM closure \cite{AndreLesieur77}.

\begin{figure}
\includegraphics[scale = 0.35]{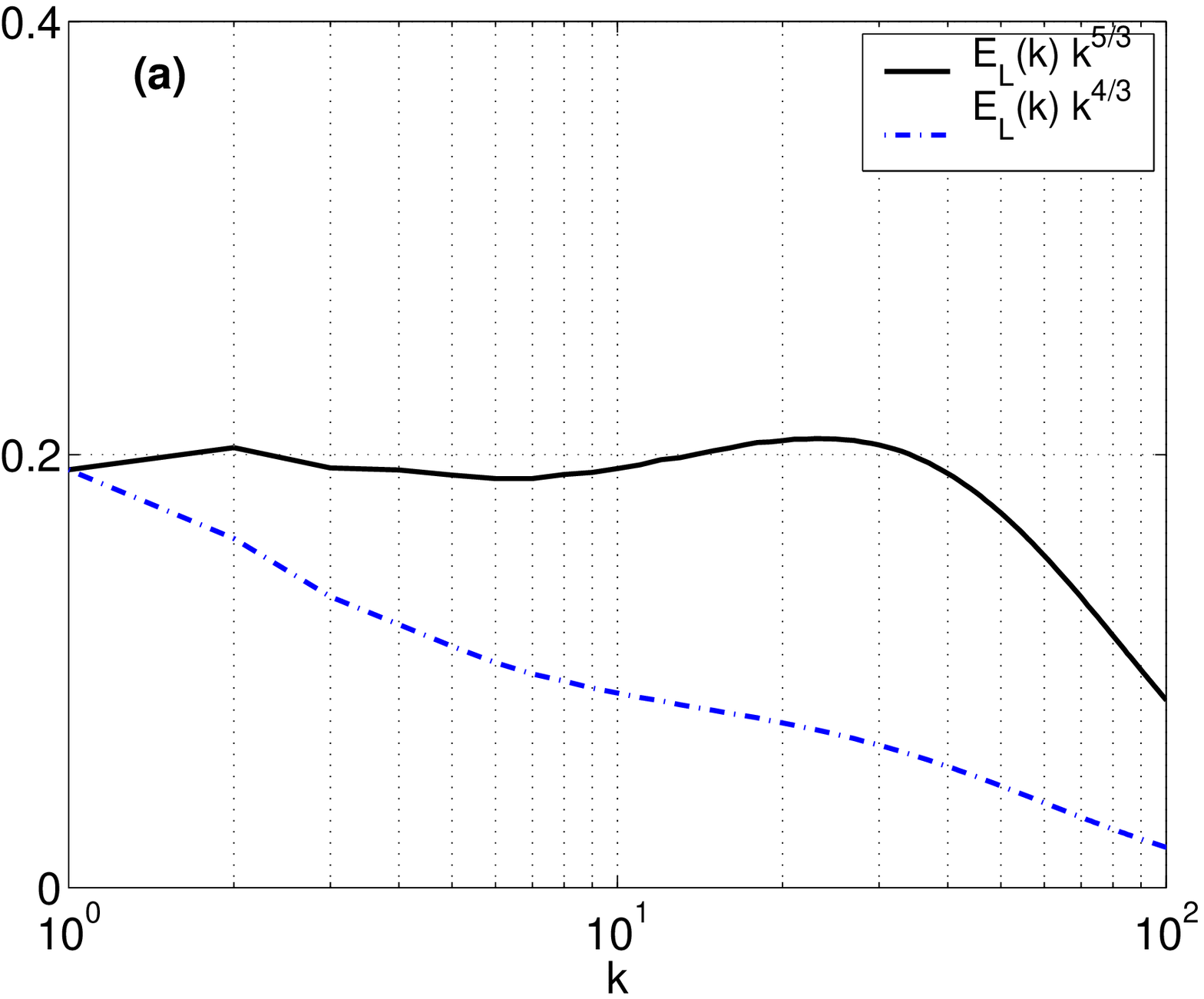}
\includegraphics[scale = 0.35]{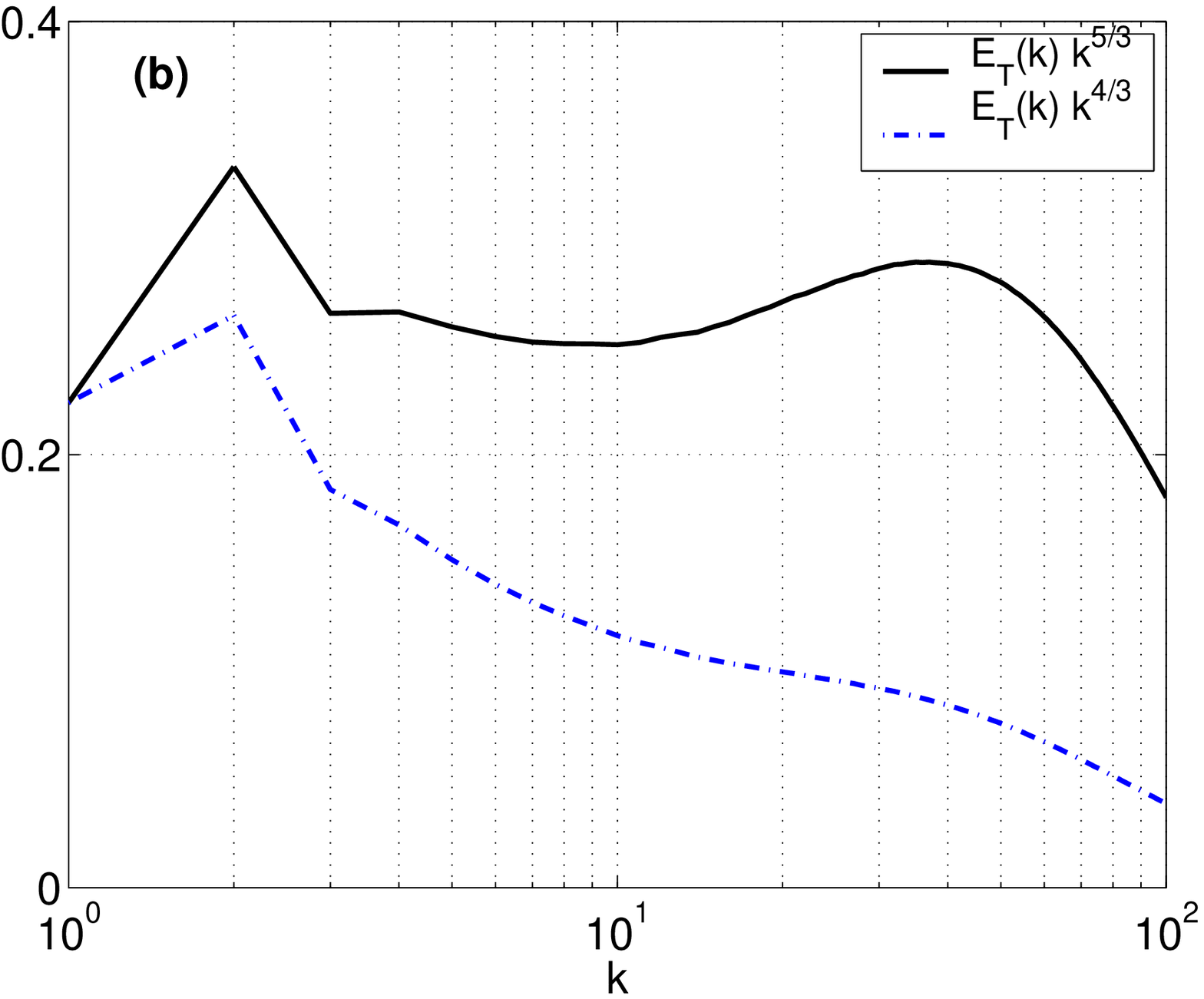}
\caption{Compensated (a) longitudinal and (b) transverse energy
  spectra for the simulation II. $E_L(k)$ shows a diminished
  bottleneck where the full spectrum $E(k)$ of Fig. \ref{spectra}(c)
  has a pronounced bottleneck.  $E_T(k)$ shows a range of $k^{-5/3}$
  coinciding with that of Fig \ref{spectra}(c) and a persistent
  bottleneck with a scaling shallower than $k^{-5/3}$ scaling in the region of
  the bottleneck.\label{spec1-d}}
\end{figure}

Figure~\ref{spec1-d} shows the compensated longitudinal and transverse
spectra for simulation II. The bottleneck is greatly diminished in
$E_L(k)$ (Fig.~\ref{spec1-d}(a), solid curve) which shows close to
$k^{-5/3}$ scaling throughout the inertial range. In $E_T(k)$ (Fig.
\ref{spec1-d}(b), solid curve), the bottleneck persists although its
peak occurs slightly earlier in wavenumber than the bottleneck for the
corresponding total spectrum. For completeness, we have also shown the
1-$d$ spectra compensated by $k^{4/3}$ (Fig.  \ref{spec1-d}, dotted
curves). Based on our arguments above, we might have expected a
stronger $k^{-4/3}$ scaling region of the transverse spectrum; such
behavior is not clearly observed although there is a tendency towards
a scaling shallower than $k^{-5/3}$ in the bottleneck regime. We plan
in future work to check the present indications that the bottleneck
will be stronger in the transverse spectrum than in the longitudinal
one because of the greater contamination of the former by the helical
co-spectrum dynamics. This raises the intriguing possibility that
$\tau_H$ affects the scaling of transverse structure functions,
accounting for some of the observed difference between the scaling
exponents of longitudinal and transverse structure functions in
near-isotropic, high-Reynolds number turbulence data \cite{SreDhr98}.
Such contributions would appear as a parity-breaking in the
$isotropic$ small-scales, and might not be easily disentangled by, for
example, the SO(3) group decomposition methods (see \cite{KurSre_LH01}
and references therein) used to extract isotropic contributions to
non-helical turbulence statistics. Although this aspect of the
influence of helicity dynamics remains speculative we hope this work
provides sufficient motivation for further investigation.

The bottleneck is a well-known phenomenon which has been observed in
experimental measurements \cite{SadVee94,KRS95} and Navier-Stokes
simulations \cite{MCDKWZ97,Gotoh02}.  While the mean helicity of the
flows in these investigations might be zero (although many do not
report the mean helicity, using the often reasonable assumption that
the flow is non-helical on average), their relative helicities might
not be, and indeed have not been reported, because the connection
between the Kolmogorov phenomenology, helicity dynamics and the
bottleneck did not exist. There have been various different approaches
taken to explain this phenomenon, including viscous effects
\cite{Falkovich94} and various kinematic arguments used to fit to a
parametrized form \cite{SheJac93,LohMul95,LohMul96}.  To the best of
our knowledge, there has thus far never been an association of the
bottleneck with helicity. What we are proposing here is a fundamental
physical cause of the bottleneck due to the helicity dynamics slowing
down the cascade of energy and helicity, the two conserved quantities
in turbulence. Further, our empirical evidence, particularly in case
II, where the total helicity is negligibly small, indicates that this
effect could occur even in flows with essentially zero mean helicity
but with non-zero relative helicity spectra (i.e. $H(k)$ is $not$ zero
everywhere).
Said differently, even if $\int H(k) dk = 0$, there can be a range of
$k$ where $|H(k)|$ and hence $|H(k)|/2kE(k)$ is finite and possibly 
large enough that $\tau_E/\tau_H \sim (|H(k)|/2kE(k))^{1/2}$ is
not negligible. This effect was reported in an experimental work
investigating spontaneous reflection-symmetry breaking in boundary
layer flows \cite{KhoShaTsi01}. Such a
scenario may occur since, while global helicity is statistically
conserved in the inertial range, local helicity is not.  Let $\xi({\bf
  x})$ be the local helicity $\xi({\bf x}) = {\bm u}({\bf x})\cdot{\bm
  \omega}({\bf x})$. The equation for $\xi$ reads
\begin{eqnarray}
\partial_t\, \xi + u_j \partial_j \xi &=& -\partial_j(\omega_j p)
 + \frac{1}{2}\partial_j (\omega_j |{\bm u}|^2)\nonumber\\
&&  + \nu [\, \nabla^2 \xi - 2(\partial_i u_j)(\partial_i \omega_j) \,].
\end{eqnarray}
Locally, both the nonlinear and
the viscous terms in the helicity dynamics might play a role in
enhancing or diminishing the helicity. If the non-zero relative
helicity in the scaling range arises from the nonlinear term, then the
effects we see are indeed valid in the high-Reynolds number (inviscid)
limit; if they arise from the viscous term then the effects we see
would disappear at very high Reynolds numbers. This hypothesis is not
testable at the present time but our work,
which studies data from simulations which are similar to, or the same
as, several performed before (see for example \cite{SVBSCC96,
  GotFukNak02, TayKurEyi03}), with comparable Reynolds number, 
gives substantial motivation to further examine these questions.
We hope in particular to motivate measurements of the relative
helicity and scaling behavior of the bottleneck region, in
other flows which report the bottleneck phenomenon in order to further
check this connection.

We finally present a key result from analysis of the convergence of
the dissipation integrals for a two-timescale cascade. In \cite{CCE03}
the same analysis was performed assuming a single timescale $\tau_E$.
The total dissipation of energy and helicity may be written as 
\begin{eqnarray}
D_{E}&=& 2\nu\Big(\int^{k_c}_0 dk~k^2~E(k) + \int^{k_d}_{k_c}
dk~k^2~E(k)\Big)\\ 
D_{H}&=& 2\nu\Big( \int^{k_c}_0 dk~k^2~H(k) +
\int^{k_d}_{k_c} dk~k^2~H(k)\Big),
\end{eqnarray}
where
\begin{eqnarray}
 E(k) &=& \cases{ C_E \, \varepsilon^{2/3} k^{-5/3} & for $k < k_c$;
                  \cr C_H \, \varepsilon h^{-1/3} k^{-4/3} &for $k_c <
                  k < k_d$, \cr} \\ H(k) &=& \cases{ c_E \,
                  \varepsilon^{-1/3} k^{-5/3} & for $k < k_c$; \cr c_H
                  \, h^{2/3} k^{-4/3} &for $k_c < k < k_d$, \cr},
\end{eqnarray}
and $C_E, C_H, c_E$ and $c_H$ are constants.  The precise estimate of
the transition wavenumber $k_c$ is unimportant to what follows. The
upper wavenumber $k_d$ denotes the maximum beyond which the cascade is
completely suppressed by viscosity. We obtain 
\begin{eqnarray}
D_E \sim \nu~\varepsilon~ h^{-1/3}~k_d^{5/3}\\
 D_H \sim \nu~ h^{2/3} k_d^{5/3}. 
\label{kd_H}
\end{eqnarray}
 In \cite{CCE03} $k_d = k_\varepsilon \sim
(\varepsilon/\nu^3)^{1/4}$ (the Kolmogorov dissipation wavenumber). In
our case, setting $k_d = k_\varepsilon$ causes the integrals to
diverge as $ \nu^{-1/4}$ in the limit $\nu \rightarrow 0$. The choice
$k_d \sim (h/\nu^3)^{1/5}$ ensures that the integrals converge to
their correct values for statistically steady-state, $D_E
=\varepsilon$ and $D_H = h$. We will call this new wavenumber $k_h$
since it depends solely on the helicity dissipation rate. It must be
distinguished from the $k_H$ of \cite{DitGiu01} which depends on both
energy and helicity dissipation rates. In the limit $\nu \rightarrow
0$, $k_\varepsilon \gg k_h$. In our simulations $k_\varepsilon > k_h$
by a factor of about 2.5. While in agreement with our analysis, we
cannot really distinguish between the two wavenumbers in this data.
However, we suggest that the resolution requirement for
measurements in turbulence with helicity, or more precisely, with non-zero helicity spectra, might be weaker than that in
turbulence without helicity.

\begin{acknowledgments}
  We are grateful to U.~Frisch and K.~Ohkitani for useful discussions
  and for bringing our attention to Refs.~\cite{DHYB03} and
  \cite{ConMaj88} respectively. Part of this work was completed during
  the visit of two of the authors (SK and TM) at the Observatoire de
  la C\^ote d'Azur, Nice, France.
\end{acknowledgments}

\bibliography{heltime_pre}
\end{document}